# Ultra-clean assembly of van der Waals heterostructures

Wendong Wang[1,2*], Nicholas Clark[2,3†*], Matthew Hamer[1,2*], Amy Carl[1,2*], Endre Tovari[1,2], Sam Sullivan-Allsop[2,3], Evan Tillotson[2,3], Yunze Gao[1,2], Hugo de Latour[2,3], Francisco Selles[1,2], James Howarth[1,2], Eli G. Castanon[1,2], Mingwei Zhou[1,2], Haoyu Bai[4], Xiao Li[1,2], Astrid Weston[1,2], Kenji Watanabe[5], Takashi Taniguchi[5], Cecilia Mattevi[4], Thomas H. Bointon[2], Paul V. Wiper[7], Andrew J. Strudwick[7], Leonid A. Ponomarenko[1,6], Andrey Kretinin[1,2,3], Sarah J. Haigh[2,3], Alex Summerfield[2†] and Roman Gorbachev[1,2,8†]

* These authors contributed equally to this work.
† Corresponding authors, emails:
roman@manchester.ac.uk
nick.clark@manchester.ac.uk
alex.summerfield@manchester.ac.uk

**Affiliations:**
[1]Department of Physics and Astronomy, University of Manchester; Manchester M13 9PL, UK
[2]National Graphene Institute, University of Manchester; Manchester M13 9PL, UK
[3]Department of Materials, University of Manchester; Manchester M13 9PL, UK
[4]Materials Department, Imperial College London, London, SW7 2AZ, UK
[5]National Institute for Materials Science, Namiki 1-1 Tsukuba, Ibaraki, 305-0044, Japan
[6]Department of Physics, University of Lancaster; Lancaster LA1 4YW, UK
[7]Graphene Engineering Innovation Centre; University of Manchester, Manchester M13 9PL, UK
[8]Henry Royce Institute for Advanced Materials; University of Manchester, Manchester M13 9PL, UK

**Abstract:** Layer-by-layer assembly of van der Waals (vdW) heterostructures underpins new discoveries in solid state physics, material science and chemistry. Despite the successes, all current 2D material (2DM) transfer techniques rely on the use of polymers which limit the cleanliness, ultimate electronic performance, and potential for optoelectronic applications of the heterostructures. In this article, we present a novel polymer-free platform for rapid and facile heterostructure assembly which utilises re-usable flexible silicon nitride membranes. We demonstrate that this allows fast and reproducible production of 2D heterostructures using both exfoliated and CVD-grown materials with perfect interfaces free from interlayer contamination and correspondingly excellent electronic behaviour, limited only by the size and intrinsic quality of the crystals used. Furthermore, removing the need for polymeric carriers allows new possibilities for vdW heterostructure fabrication: assembly at high temperatures up to 600°C, and in different environments including ultra-high vacuum (UHV) and when the materials are fully submerged in liquids. We demonstrate UHV heterostructure assembly for the first time, and show the reliable creation of graphene moiré superlattices with more than an order of magnitude improvement in their structural homogeneity. We believe that broad adaptation of our novel inorganic 2D materials assembly strategy will allow realisation of the full potential of vdW heterostructures as a platform for new physics and advanced optoelectronic technologies.

**Introduction:**

Van der Waals heterostructures are a versatile platform for creating unique quantum and optoelectronic metamaterials with atomically sharp and clean interfaces[1,2]. Over the last decade, hundreds of proof-of-principle optoelectronic devices with novel functionalities and exciting physical properties have been demonstrated, based both on individual atomically thin crystals and their heterostructures. The fabrication of such artificial solids generally relies on the mechanical stacking of individual two-dimensional materials (2DM). To date, a wide variety of polymer-assisted transfer approaches exist[3], which can be generally categorised in two groups: **1.** where 2DMs come in direct contact with liquid, often referred to as "wet"[4,5] and **2.** where the surface of 2DMs remain dry throughout the transfer process, thus termed "dry" transfer[6]. These techniques are employed to pick up grown[4,7–9] or exfoliated 2D crystals[5], and can be used successively to assemble multi-layer stacks[10,11]. For all approaches used to date a polymer support is used to manipulate the 2DM or vdW heterostructure, creating the stack and depositing it onto a substrate compatible with the desired function of the final device.

Achieving 2D heterostructures that behave as artificial crystals, rather than independent materials, requires atomically clean interfaces; transistors and LEDs fabricated from vdW heterostructures only operate in those specimen areas free from contamination[3]. Furthermore, many of the most challenging fundamental physics experiments are achieved only when charge carriers exhibit ballistic transport[12], behaviour which is only seen in the limit of extreme cleanliness. Unfortunately, the necessity to use polymers in the 2D transfer and stacking process provides a key source of interlayer contamination. The polymer layers provide excellent adhesion and flexibility, but their residues remain at buried interfaces and/or on the surface of the completed heterostructures (depending on the method used)[3]. A second source of undesirable surface and interfacial contamination comes from the surrounding environment. High and ultra-high vacuum (HV/UHV) processing would provide a route to remove the potential for volatile species to condense on surfaces during the stacking process but it is not compatible with the polymer transfer approaches (due to degassing of the polymer). Consequently, to date, all vdW heterostructures have been assembled in air, inert gases[13,14] or low vacuum[15]. The strong adhesion between 2DM causes contaminants to aggregate into localized pockets known as blisters[16] or bubbles[8,12] consisting of air[16], water[17], and hydrocarbons[2]. This localized contamination limits the size of homogeneous regions with atomically clean interfaces[18] to micrometres, severely restricting the experiments that can be performed and hampering the development of advanced electronic applications where large uniform areas are required for dense and reproducible patterning of circuit elements. For graphene and hexagonal boron nitride (hBN) heterostructures, performing transfers at elevated temperatures (e.g. 110°C) and gradually collapsing the contact front of the polymer (i.e. "slow transfer")[16] has shown to promote the diffusion of surface contaminants, preventing them from being trapped at the interface as it is formed. The use of higher temperatures could facilitate the formation of larger areas with atomically clean interfaces for a wider range of 2DMs but the use of a polymer carrier limits the permitted temperature to <150°C (varies with the glass transition temperature of the chosen polymer). Besides this, the presence of polymers also prevents the use of other methods to achieve high cleanliness such as aggressive cleaning techniques (e.g. most organic solvents swell/dissolve typically used polymers). Strategies to remove the contamination from already assembled heterostructures exist[18,19], however the cleaning process is slow even for micrometre-scale devices, cannot be widely adopted, and risks damaging the structure due to the excessive heat and/or high mechanical stress involved.

Lastly, some of the most exciting recent advances in 2DM have been centred on crystals that experience chemical degradation during exfoliation and transfer in ambient environmental conditions[3,20]. This problem is often mitigated by operation in pure argon or nitrogen atmospheres[13,14]. However, when atomically thin, many of the most interesting crystals such as $CrI_3$ and $InSe$[21] still degrade when processed in a state-of-the-art inert glove box environment due to the residual oxygen and water. Developing a transfer method compatible with UHV is therefore essential to enable the characterisation and exploitation of the unique electronic, optical, and magnetic properties of atmospherically sensitive next generation 2DMs.

In this work we present a new technology platform for fabricating ultra-clean 2D material heterostructures which entirely avoids the use of any organic materials. This approach allows far greater flexibility in the transfer conditions than has been possible to date, including the first assembly in ultra-high vacuum (UHV) at a pressure of $10^{-10}$ mbar. To replace the polymers, we employ inorganic flexible $SiN_x$ membranes covered with thin metallic films. The film thickness and composition can be changed to allow precise tuning of the adhesion properties, enabling high transfer reliability with nearly 100% yield for a wide range of 2DMs. We exemplify this technique by performing transfers in air, glove box and UHV environments at temperatures up to 350°C, enabling reliable fabrication of devices with clean areas limited only by the size of the crystals used. All studied graphene devices have high quality and demonstrate carrier mobilities higher than $1.0 \times 10^6$ cm$^2$V$^{-1}$s$^{-1}$ at 5K. In addition, we demonstrate dramatic increases in the atomically clean areas of complex multilayer heterostructures, and moiré uniformity in twisted devices, as well as demonstrate scalable transfer of CVD grown materials. We envisage this technique becoming a new standard in 2DM nanofabrication, removing the cleanliness 'bottlenecks' which have been plaguing researchers for the last decade, as well as allowing for alternative scale-up pathways[22,23].

**Inorganic assembly in air/glove box**

The transfer relies on chemically inert, flexible and transparent $SiN_x$ membranes. For smaller heterostructures built from mechanically exfoliated crystals with lateral dimensions of ~10s micrometres, the membranes are typically shaped into a cantilevers, each 320-480 μm long by 160 μm wide protruding from a silicon chip (Fig.1a). These are fabricated at wafer-scale using commercially available silicon wafers coated with 500 nm low pressure CVD-grown $SiN_x$ on both sides. Optical lithography and reactive ion etching (RIE) are used to pattern the nitride layers into the required cantilever geometry before wet etching is employed to selectively remove the underlying silicon and release the cantilevers (see Supplementary Section 1).

The strong adhesion between 2DMs and polymer membranes is the key factor that has made polymer-based transfer methods universal. In comparison, bare $SiN_x$ presents poor adhesion to 2DMs. To address this problem, we coat the cantilevers with a tri-metal stack consisting of 1 nm Ta, 5 nm Pt and 0.1-1 nm Au (Fig.1b-d). Gold generally provides strong adhesion to 2DMs[24,25] but adjusting its thickness allows us to tuning the adhesion strength for a specific 2DM/substrate combination (higher gold surface density leads to stronger adhesion). This is highly beneficial as it allows the completed heterostructures to be released onto various substrates once the desired layer combination is assembled. The role of the Pt layer is to compensate for the variable $SiN_x$ roughness and ensure consistent surface quality between different $SiN_x$ substrates. Additionally, the presence of Pt facilitates the catalytic decomposition of mobile surface hydrocarbons[26] at elevated temperatures. Finally, Ta serves as the adhesion layer for Pt. An AFM image showing the roughness of the cantilever surface is shown in Fig.1e, with typical root mean square (RMS) deviation of 130-140 pm. Energy dispersive X-ray spectroscopy (Supplementary Figure S11) shows uniform distribution of all three metals across the surface, which is not the case if the Pt layer is removed, due to the tendency of Au to form nanoclusters.

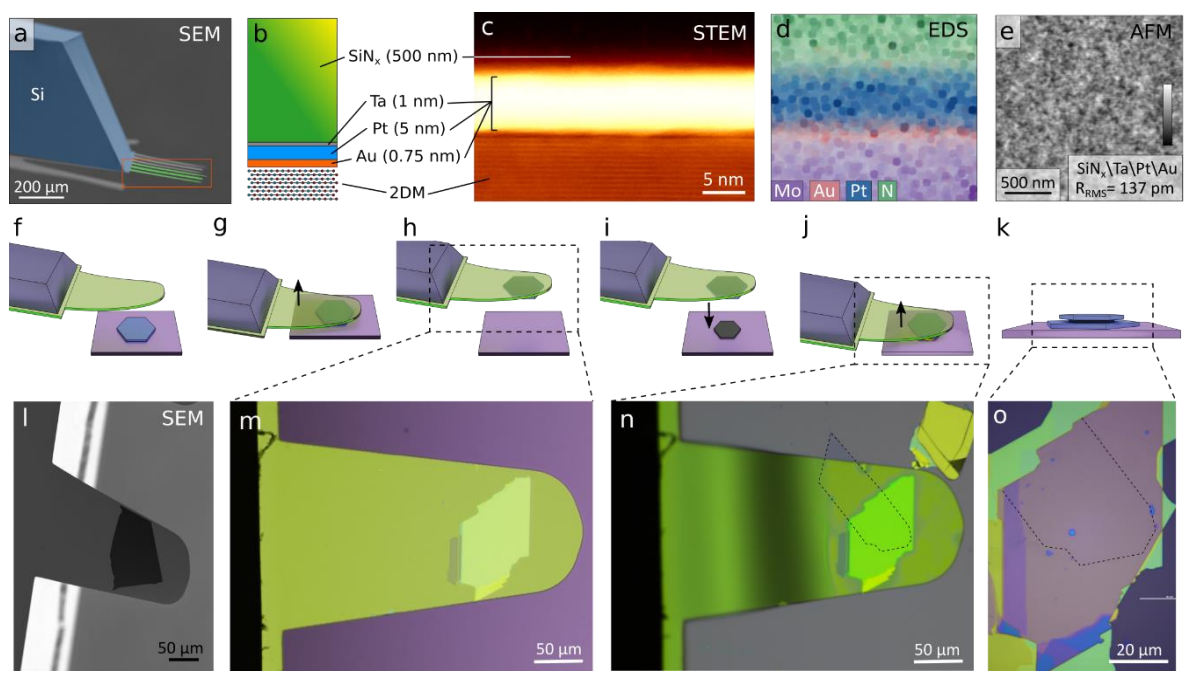

**Figure 1: Inorganic assembly of vdW heterostructures.** *(a) SEM micrograph of several cantilevers protruding from a silicon chip. (b) Schematic and (c) cross-sectional high-angle annular dark-field (HAADF) scanning transmission electron microscopy (STEM) image showing multilayer metallic coating of the cantilever holding a 2DM specimen (sample shown in a multilayer $MoS_2$ crystal). (d) Elemental mapping of the region presented in (c) using energy dispersive X-ray spectroscopy. (e) AFM micrograph of the cantilever surface after the coating process. The root mean square roughness values ($R_{RMS}$) is indicated on the image. (f-h) Steps employed to pick up an hBN crystal onto the fabricated cantilever: (f) alignment, (g) contact and (h) lift-off. SEM (l) and optical (m) micrographs of the cantilevers after pick-up of a thick (~40 nm) hBN crystals. (i, j) Steps involved in the pick-up of a graphene crystal: alignment (i), contact and lift-off (j). (n) Optical micrograph showing the cantilever in contact with the graphene (edges highlighted with dashed line) on $SiO_2$. The flexible nature of the cantilevers allows accurate control of the lamination process. (k) Graphene/hBN stack deposited onto the bottom hBN crystal. The lamination process is stopped before the entire bottom hBN crystal is covered by the cantilever to selectively release the stack instead of picking it up. (o) Optical micrograph showing the resulting heterostructure on an oxidised silicon wafer presenting large uniform areas. Further data on additional samples can be found in Supplementary Section 2.*

We benchmark the polymer-free transfer approach by demonstrating the exceptional cleanliness achievable for the archetypal hBN/graphene/hBN vertically stacked heterostructure, fabricated from mechanically exfoliated 2D crystals on an $Si/SiO_2$ substrate. As illustrated in Fig.1f, firstly the commercial 2D crystal transfer set-up (see methods) is used to align the custom cantilever over a target hBN crystal (Fig.1g) and lowered with a 20° tilt until contact is made at 120-150°C. The cantilever is raised after few seconds, peeling the crystal away from the substrate (Fig.1h). We find that the optimal gold thickness for transferring most 2DM on to $SiO_2$ is 0.65nm, providing >95% successful transfers based on over 200 operations. The cantilever with the hBN attached is then used to pick up the next layer in the heterostructure (graphene, as shown in Fig.1i,j) and the resulting stack is then deposited onto the bottom hBN by exploiting the greater adhesion to this larger bottom crystal (shown in Fig.1k). Fig.1o shows an optical micrograph of the finished stack consisting of a large clean region with no contamination or bubbles over an area larger than 25μm × 40μm, which has been formed despite a very fast (few seconds) lamination speed. The absence of polymers during the processing allows for much higher temperatures to be employed: for the stack in Fig.1 we used 150°C for the first hBN, 120°C for the graphene and 230°C for the second hBN. For transfers in air or inert gas, we have tested transfer temperatures up to 300°C and observed the complete disappearance of hydrocarbon bubbles above ~220°C. However, heating above 150°C in argon during the pick-up of monolayer graphene from $SiO_2$ may produce microcracks in the graphene, presumably due to the increase in adhesion of graphene to $SiO_2$[27] (see SI section 2 for more details). A further advantage of our approach is the improved positioning accuracy compared to polymer-based methods. We find that this is limited only by the optical resolution of the microscope used (better than 400 nm) as the inorganic $SiN_x$ membrane displays negligible drift and warping when heated and compressed (see supplementary video 1). After the transfer, membranes can be cleaned in $Ar/O_2$ plasma and reused multiple times.

To characterize the quality of the encapsulated graphene, we fabricated 4 multi-terminal Hall-bar devices with various dimensions in an argon environment, and studied them at 5K. For 3 of the samples, displaying a width ≤12.5 micrometres, we have found that the carrier mean free path ($l$) is limited by their physical dimensions, and the field-effect mobility follows the characteristic dependence $\mu = l\,(2e/h)\sqrt{\pi/n}$ (Fig. 2a) reaching well over $10^6$ cm$^{-2}$V$^{-1}$s$^{-1}$ at low carrier concentrations. This behaviour is typical for the current state of the art devices, with the hall-bar dimensions limited by to 20 micrometres when using polymer transfer[10,28,29]. However, for a larger device, with dimensions 38 x 33 μm, we do not see such saturation, in fact the carrier mobility increases with carrier density, reaching up to ≈ 3·10$^6$ cm$^2$V$^{-1}$s$^{-1}$. This behaviour suggest that a different scattering mechanism becomes relevant for large devices, probably related to carbon and oxygen substitutional impurities intrinsically present in the hBN encapsulation[30].

Despite many research articles demonstrating outstanding proof-of-principle optoelectronic performance, the random network of contamination bubbles on micrometre length-scale renders promises of such applications unrealistic. Electronic-grade materials require nearly 100% areal uniformity for the design of the dense circuit elements such as optical emission arrays or transistors, which makes even simple heterostructures such as polymer-assembled encapsulated graphene unusable for high-end electronic applications. Even more challenging is the future of applications for multi-layer heterostructures containing 2D semiconductors, such as vertical Light-Emitting Devices[11,31] typically displaying very small clean areas (<2µm × <2µm, with over 50% area covered by contamination bubbles) and taking several days or weeks to fabricate. Here, we assemble a complex structure, consisting of 8 individual 2D layers with different thicknesses ranging from bulk to a monolayer, where the optically active layers are two twisted few-layer $MoS_2$ crystals (outlined with the yellow and orange dashed lines in Fig.2b). The entire heterostructure, fabricated in an argon atmosphere, displays no bubbles, including a completely clean 9×15 µm area where all the eight layers overlap as confirmed by overall AFM and local cross-sectional STEM measurement (Fig.2c,d). A comparison of the topography to similar devices fabricated using polymer based transfer techniques in the same atmosphere is presented in Figure S6. Absence of contamination can also be seen in another LED structure (Fig.2e,f) fabricated, featuring monolayer $WS_2$ as the optically active layer, and showing an overall overlap area of 25x30 µm. Both of these assemblies took only 10 minutes to complete and resulted in over a 100-fold increase of uniform heterostructure area, limited in this instance only by the size of exfoliated crystals. When a bias voltage is applied to the graphene layers we observe a bright uniform electroluminescence signal with a terrace-like distribution of the intensity across the device (shown in Fig.2g), which is related to the change in thickness of the tunnelling hBN layers. Narrow exciton emission lines (4-5 meV) confirm the state-of-the-art optoelectronic quality of the device (Fig.2h), similar to that recently reported in small samples[32], with a slight variation in the emission energy commonly attributed to the non-uniform impurity distribution in the original crystal[33]. This exemplifies the importance of our technique, as the advanced optical performance previously only observed on selected specimen locations ~2 µm large can now be realistically considered in the context of advanced applications.

The developed transfer process was also tested for a variety of air sensitive 2D materials that require fabrication in an argon environment (e.g. black phosphorus, $WSe_2$, $NbSe_2$) exfoliated on different substrates, including $SiO_2$ and polymers such as PMMA and PPC spin-coated onto Si wafers. In all the cases we observe complete absence or considerable reduction in the presence of hydrocarbon pockets, see Supplementary Fig.S5a-c. Lithographically patterned 2D crystals can also be transferred (Fig.S5d-e), however one needs to be aware that some dry etching processes lead to chemical bonding of 2D crystals to the substrate's surface along the etched perimeter.

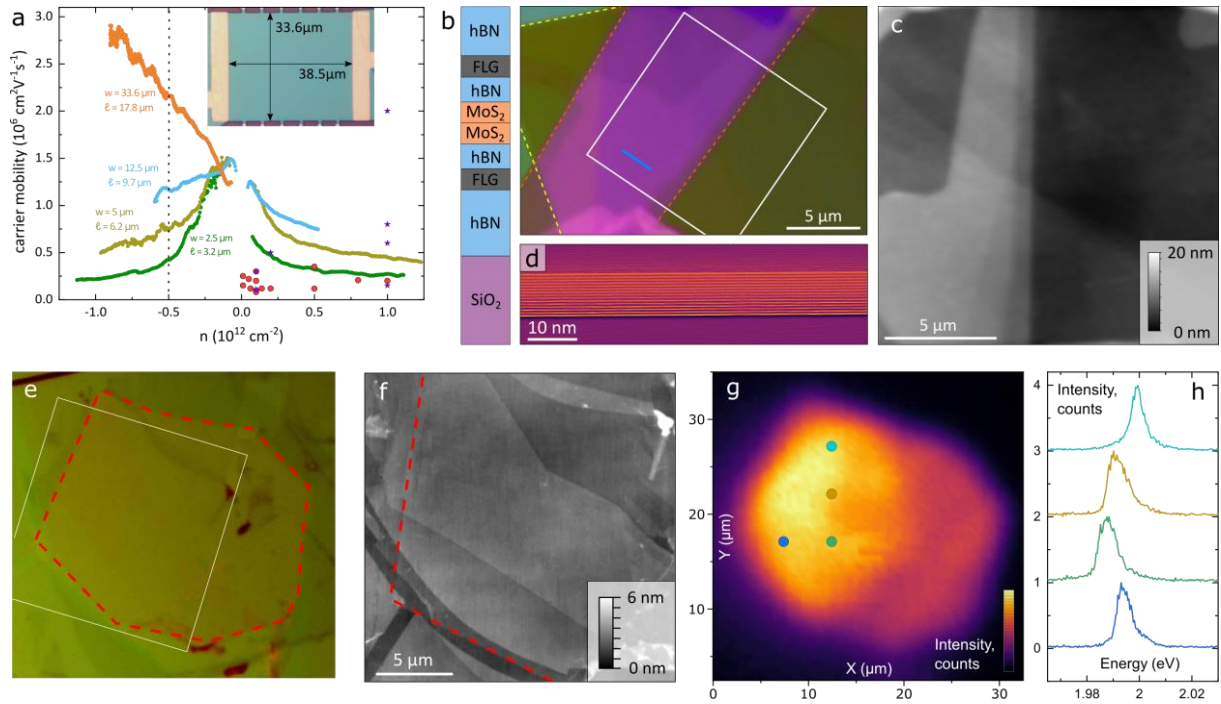

*Figure 2: Heterostructures assembled using our fully inorganic transfer technique in air and argon. (a) Carrier mobility at 4 K for 4 graphene devices. For large devices (>20 µm) the mobility is no longer limited by the edge scattering and increases with carrier density. For smaller devices, the extracted values of l agree well with the dimensions of the studied devices. Points indicate mobility values (calculated using Drude relation) acquired from published literature where polymeric transfer has been used. Purple circles correspond to samples fabricated at Manchester University, and red to other groups (see Supplementary Table 1 for full breakdown). (b) Optical and (c) AFM micrographs of an 8-layer stack assembled in air. No bubbles are present, although some crystals are folded during the pick-up process. (d) Cross-sectional high-angle annular dark-field (HAADF) scanning transmission electron microscopy (STEM) image showing cross-section of the device at the location marked with blue line in (b), graphene and hBN layers are indistinguishable due to the close atomic mass values. Local contrast enhancement has been applied to the image to highlight the layered structure. (e, f) optical and AFM micrographs of an LED-type structure[11] with monolayer $WS_2$ selected as the optically active layer. (g) Electroluminescence intensity map for the device shown in (e) showing a highly uniform emission distribution with a pronounceable terraced structure due to variation in the hBN barrier thickness (3-5 layers). (h) Example of electroluminescence spectra acquired on the points labelled in (g).*

Previous work has shown improved heterostructure device performance when the transfer is performed in organic solvents to reduce wrinkling and strain[34,35]. However, these experiments were limited to polymer compatible solvents (alcohols, alkanes). In contrast, our fully inorganic transfer method allows transfer when submerged in almost any organic solvent, for example ketones or chloroform, which are more likely to effectively remove contamination[27,36] during the transfer. Indeed, we have confirmed the transfer also works in a variety of organic solvents incompatible with the polymer-based transfer method (toluene, chlorobenzene, cyclohexane, chloroform, various ketones, etc), although in some cases (mostly for polar solvents) a second micromanipulator was required to push the cantilever down to touch the substrate, presumably due to the formation of a repulsive solvent shell on the solid surfaces. Example of this process performed in acetone can be found in Supplementary Figure S5.

**Ultra-high vacuum assembly**

Further improvements in vdW cleanliness are impossible in the presence of atmospheric contamination.[37] To eliminate the interlayer contamination completely, we designed and built a 15-axis micromanipulation setup enclosed within a UHV chamber, with a base pressure of $4\times10^{-10}$ mbar, Fig.3a. The heterostructure assembly followed the procedure described in Fig.1, using a custom-built UHV optical microscope integrated into the UHV chamber. Further details of the UHV system design can be found in Supplementary Section 3.

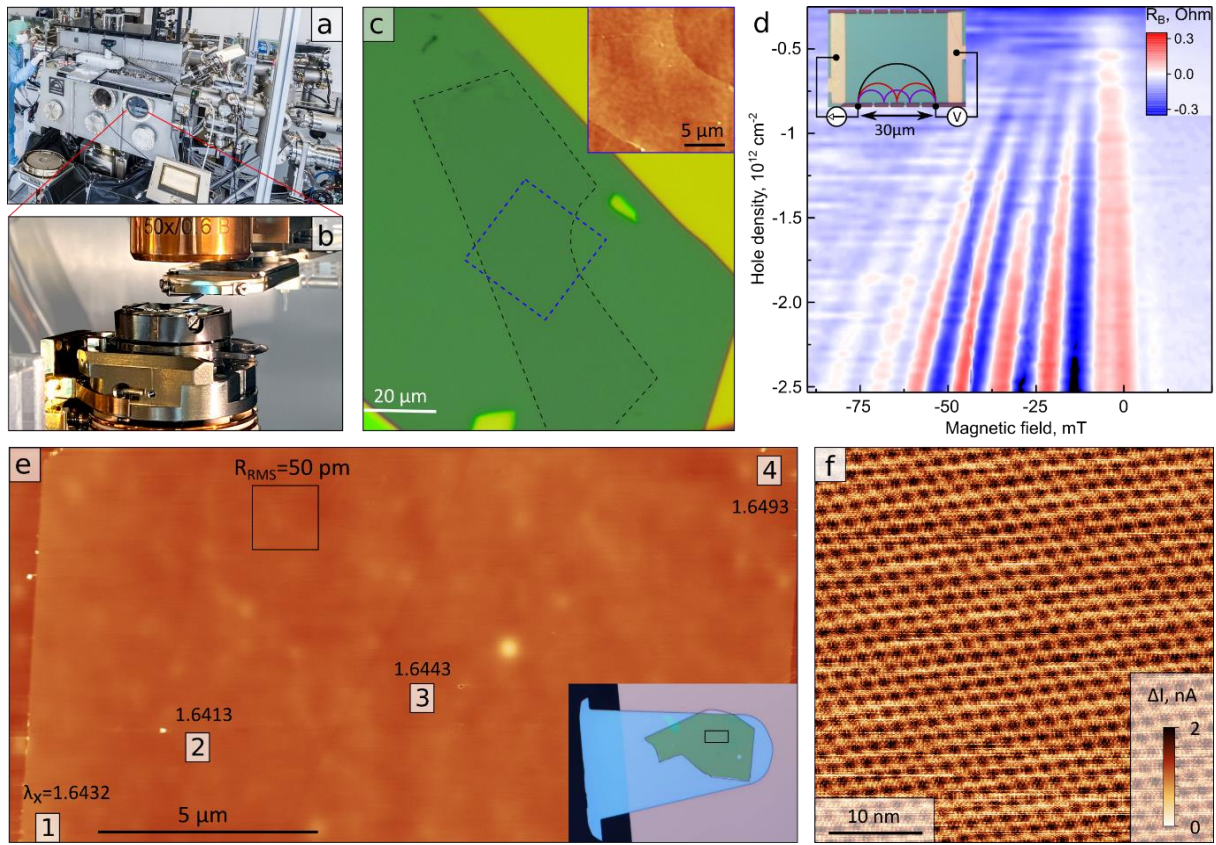

*Figure 3: Ultra-High Vacuum assembly of van der Waals heterostructures. (a) Instrument for performing UHV assembly (containing a 15-axis manipulation 2DM transfer set-up and an integral optical microscope). (b) Close-up of the optical lens, cantilever and sample stage. (c) Example of a UHV-fabricated vdW heterostructure: hBN – encapsulated graphene on graphite (serving as a local back-gate). Graphene is outlined with a black dashed line. No hydrocarbon contamination is observed in the entire sample. Inset shows a 25 μm AFM scan of the area highlighted by a blue-dashed box. (d) Magnetic focusing experiment performed at 5K on a similar heterostructure as in (c), indicating ballistic transport on the scale of >30 micrometres, corresponding to a carrier mobility over $6×10^6$ $cm^{-2}V^{-1}s^{-1}$. Inset shows the measurement schematic: the electrons injected from one contact are magnetically deflected into the other contact (collector) while the corresponding voltage is measured. The transfer resistance displays multiple focusing peaks, with the first peak corresponding to the direct transfer of electrons from the injector to the collector, and for the subsequent peaks (at higher magnetic field) the electrons scatter from the edge of the device before reaching the collector. (e) AFM topography map of a twisted graphene monolayer/bilayer heterostructure placed on a bulk hBN crystal. Numbers provide the x-component of the moiré period extracted at indicated locations using cAFM measurements. Small white bead-like objects are gold clusters resulting from contact deposition using a stencil mask, and the elevated areas are protruding roughness of the SiNx support. Inset shows the optical micrograph of the stack on the cantilever broken on the edge of a silicon wafer. (f) example of cAFM micrograph collected on sample shown in (e) used for moiré period analysis.*

The UHV fabricated vdW heterostructures do not display contamination bubbles across the entire interface (Fig.3c), regardless of the transfer speed and crystals' dimensions. Electronic transport measurements on an exemplar UHV fabricated vdW heterostructure, comprising hBN/graphene/hBN/graphite, shows ballistic transport over a scale of ~40μm micrometres, seen via magnetic focusing experiments[38] (Fig.3d), and a field-effect mobility of ~$3\times10^6$ $cm^{-2}V^{-1}s^{-1}$ (at n=$0.5\times10^{12}cm^{-2}$). To achieve this, all 2D crystals on Si/SiO$_2$ substrates introduced to the chamber went through UHV annealing at 400°C and were assembled into heterostructures at 150°C (Fig.3b). We have found that higher conditioning temperatures (600°C) result in a noticeable decrease in the low-temperature mobility of graphene (μ<$4\times10^5$ $cm^{-2}V^{-1}s^{-1}$ in four samples studied). Although the hBN-encapsulated graphene is immune to degradation at high temperatures[39], annealing on the Si/SiO$_2$ substrate can likely lead to local damage to the graphene lattice.

This extreme cleanliness is only possible when polymer-free transfer and UHV conditions are combined. To verify this, we have fabricated similar hBN/graphene/hBN heterostructures in UHV, but using the traditional transfer technique with small hemispherical PMMA droplet deposited on a glass slide instead of the SiNx cantilever (see Supplementary Section 5). Despite using an analogous annealing protocol, we have found a considerable number of hydrocarbon pockets in the heterostructures. This indicates that the use of polymer makes contamination unavoidable even in the UHV environment since the loose polymer matrix can emit hydrocarbons after a prolonged temperature treatment and pumping.

In addition, we demonstrate the applicability of the fully inorganic transfer approach in UHV environment to producing twisted vdW heterostructures, where perfect cleanliness leads to exceptional uniformity, opening opportunities for new experimental studies. The ability to exploit a small local twist between neighbouring 2DM to generate exquisite physics is one of the degrees of freedom not seen in any other realm of synthesis. Yet the understanding and exploitation of the properties of such structures is held back by the poor specimen uniformity, with micrometre-sized areas often displaying local variations in twist angle up to 0.2° , resulting from wrinkles present due to contamination bubbles[40]. In twisted bilayer graphene, this variation leads to difficulties in studying strongly correlated phenomena such as superconductivity[41] and magnetic states[42], which are extremely sensitive to the twist angle disorder. We have used our fully inorganic UHV assembly to fabricate a twisted monolayer/bilayer graphene on a thick hBN substrate.

AFM topography of the fabricated twisted bilayer reveals no contamination bubbles over the entirety of the heterostructure (Fig.3e). Additionally, conductive AFM (cAFM) measurements reveal that the moiré period of the superlattice is exceptionally uniform (Fig.3f) with the extracted twist angle $\theta \approx 8.5°$, varying by just 0.016° over a length scale of 10 μm (areas labelled 1,2,3), and within 0.04° for the entire area shown. This result represents an order of magnitude improvement compared to the current twist angle disorder achieved with polymer-based transfer techniques.

**Scalability**

The presented cantilever geometry can also be employed for transfer of CVD-grown materials, as exemplified in Supplementary Fig.S4, where $WS_2$ monolayer crystals[43] have been picked up from their growth substrate ($SiO_2$) and encapsulated in hBN. However, the free-standing nature of the 500 nm thick $SiN_x$ makes it difficult to apply this technique over lateral scales larger than ~200 μm. To overcome this problem, we have laminated 20x20 mm metal-coated $SiN_x$ membrane onto a 1 mm thick PDMS film, Fig.4a. Direct contact of 2D crystals with PDMS surface is known to produce surface contamination[44] due to the presence of uncrosslinked oligomers within the PDMS bulk. However, the SiNx membrane provides an impermeable barrier for this contamination, allowing us to combine the best of both worlds: the mechanical flexibility and support of PDMS with the ultraclean interface of the metallized $SiN_x$. This arrangement has allowed us to pick up large CVD-grown $WS_2$ sheet from its growth substrate ($SiO_2$) and deposit it onto a ~6L thick $WS_2$ film grown on sapphire substrate. The resulting heterostructure, shown in Fig.4b, shows 0.7±0.1 nm AFM step-height of the top layer suggesting absence of contamination in the interface and consistent with the 50-60% reduction in the PL intensity in the overlapping area (Fig.4d). Despite the use of PDMS, we are still able to perform transfers at 250°C, though other grades of PDMS allow for higher working temperatures (e.g. 350 °C[45]).

However, we found that the roughness of the growth substrate plays an important role: while the process works for flat substrates displaying few nm RMS (e.g. $SiO_2$/sapphire), we had difficulties picking-up commercially produced graphene grown on copper foil as its roughness reaches micrometre scale, thus preventing the uniform contact of the SiN$_x$ membrane with graphene. However, recent developments in growth of ultraflat metallic surfaces and their use as substrates for CVD growth of 2DMs [46–48], optimised growth processes for increased large scale uniformity[49,50], and improvements in direct growth on flat non-metallic substrates[51], should allow for use of inorganic transfer when the materials are made widely available.

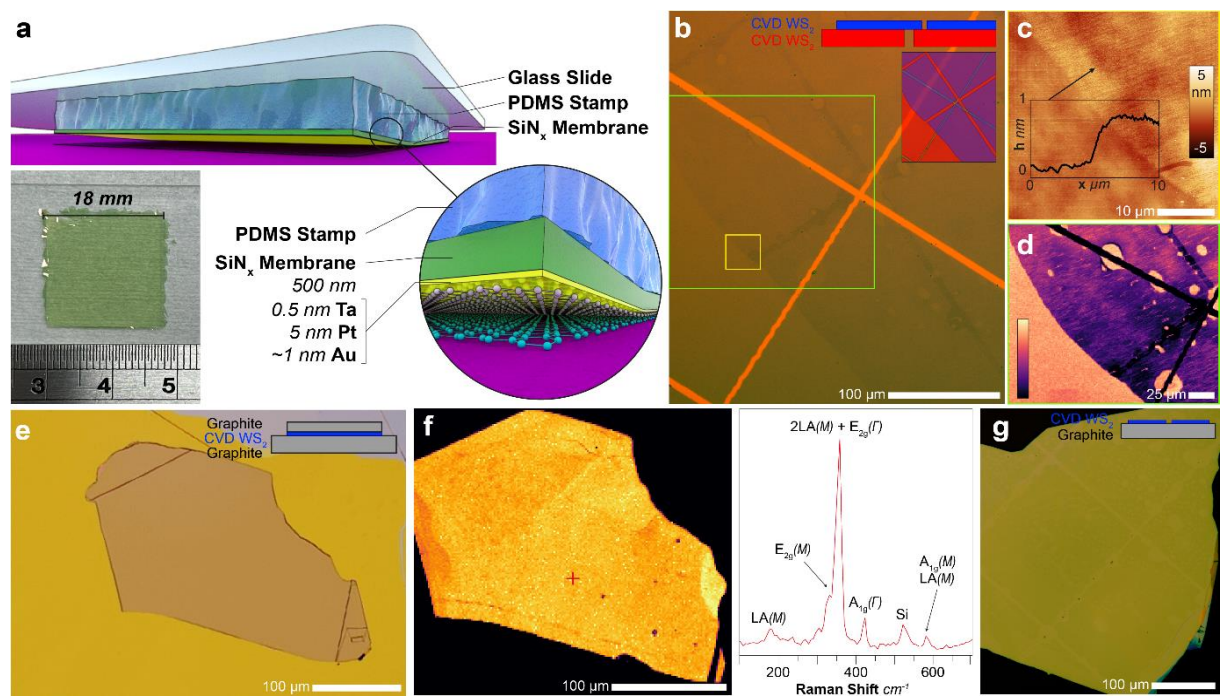

*Figure 4: Large area transfer of CVD grown 2D materials. (a) schematic showing arrangement of PDMS polymer-supported SiN$_x$ membrane. Inset image shows an 18mm SiN$_x$ membrane laminated onto PDMS film. (b) A heterostructure consisting of CVD-grown monolayer WS$_2$ transferred onto CVD-grown few-layer WS$_2$ on sapphire substrate fabricated using the laminates shown in (a). A square grid has been scratched into both layers prior to transfer to enable visualisation. The inset shows the two layers in red and blue to highlight the covered areas. (c) Topography of the area in (b) indicated by the yellow rectangle. A height profile at the indicated position is shown in the inset, measuring a step of approximately 0.7 nm. (d) Integrated intensity map of the primary WS$_2$ photoluminescence peak around 1.97 eV. (e) A thin WS$_2$ layer encapsulated in thick graphite fabricated using the large area stamping method. (f) Map of the intensity of the WS$_2$ 2LA/E$_{2g}$ Raman peak normalised by the Si peak and a representative Raman spectrum from the location indicated by the red cross. (g) A monolayer WS$_2$ CVD grown layer transferred onto a mechanically exfoliated graphite crystal. A square pattern has been scratched into the WS2 monolayer to enable visualisation.*

**Conclusion and future outlook**

The presented fully inorganic 2D transfer technology platform enables a significant improvement in the cleanliness, size uniformity, and quality of vdW heterostructures assembled in air and inert gasses. Furthermore, it allows complete elimination of all interlayer contamination when used in a UHV environment. This development provides access to highly uniform vdW heterostructure devices where the performance is limited only by the intrinsic quality and dimensions of the 2D crystals, which will to lead to new research avenues for uncovering novel scientific phenomena.

The $SiN_x$ films are readily commercially available at wafer scale, and possess superior chemical, thermal and mechanical stability compared to organic polymers. Their use as transfer membranes will significantly improve the chances of successful development of 2D heterostructures for advanced electronic technologies. While the free-standing "cantilevers" are more suitable for small scale research applications, the technology developed here relies on the general properties of the $SiN_x$/metal interface and can be used in large area laminates, providing a path to wafer-scale transfer. In addition, compatible alternative pathways to scale up have been developed using autonomous robotic searching and assembly of two-dimensional crystals[22,23]. We envisage that the outstanding reproducibility, cleanliness and speed offered by our technique, combined with machine learning algorithms to guide[52] the decision-making process, can lead to small scale manufacturing of custom high-end optoelectronic devices for applications where large batch production is not required, such as quantum technologies, aerospace and healthcare.

**Methods:**

**SiNx cantilever fabrication:** $SiN_x$ transfer cantilevers were fabricated from silicon wafers coated on both sides with low stress (non-stoichiometric) $SiN_x$ films, purchased from Inseto UK. The nitride layers on either side were patterned using optical lithography and subsequent reactive ion etching to define both the cantilevers and a hard mask for anisotropic etching of the Si. KOH solution was then used to remove the silicon from the area defined by the $SiN_x$ mask. After cleaning, the metallic adhesion layer was deposited using direct sequential sputtering of Ta and Pt, followed by e-beam deposition of Au immediately before use. See Supplementary Section 1 for details.

**2D heterostructure Fabrication:** Crystals were mechanically exfoliated onto oxidized silicon wafers. For assembly in air or argon atmosphere, cantilevers were used to pick-up/drop-off 2D crystals using a standard micromanipulator system[3] with a temperature controlled stage. For UHV assembly, exfoliated crystals were loaded into a custom UHV system and annealed in a load-lock (pressure ~$10^{-7}$ mbar) prior to introduction into the main UHV chamber (pressure ~$10^{-10}$ mbar) to remove atmospheric hydrocarbon contamination and adsorbed $H_2O$. Separately, a metal-coated $SiN_x$ cantilever was loaded and annealed prior to introduction to the main chamber via the same load-lock. After annealing, cantilevers/substrates were loaded into the top/bottom precision manipulators respectively, using a combination of linear and 'wobble-stick' style UHV manipulator arms. For stacking operations, the manipulators were controlled using a custom LabView-based program with the substrate/stamp positioned underneath an array of UHV objective lenses (5/20/50/100 X magnification) for inspection and precise alignment of the 2D crystals. Cantilever and 2D crystals were brought into contact at an elevated substrate temperature, ~150°C, to form the heterostructure stacks. Between each stacking step the partially assembled heterostructure was annealed to remove surface contaminants. Full technical details of the UHV transfer procedure and system, together with the annealing recipes for each step can be found in Supplementary Sections 3-5.

**Transport Device Fabrication:** After assembly, hBN/graphene/hBN stacks were dropped-off from the cantilever onto a large graphite crystal (50-100nm thick) pre-exfoliated onto a $Si/SiO_2$ wafer and used for applying the back gate. The hBN/graphene/hBN/graphite heterostructures were progressed into standard Hall-bar devices using e-beam lithography and reactive ion etching. Electrical contacts consisting of 3nm Cr and 70nm Au were made to graphene via the one-dimensional technique[8].

**UHV cAFM device Fabrication:** Twisted graphene samples for cAFM were fabricated using the tear and stack method [53]. Firstly, a large hBN crystal was picked up on a cantilever. Graphite was mechanically exfoliated onto an oxidised silicon wafer and monolayer / bilayer graphene crystals were identified using optical contrast. The hBN crystal on the cantilever was lined up with the edge of the bilayer section of the graphene crystal and slowly brought into contact with the surface. The cantilever was then picked up ripping the graphene along the edge of the hBN crystal. The section of graphene crystal remaining on the substrate was then rotated to the desired angle, realigned with the first half already on the cantilever, and picked up. The cantilever with the hBN/2Lgr/gr heterostructure was then flipped upside down and broken off on the edge of a $SiO_2$ wafer coated with a metallic adhesion layer (2nm Ti, 20nm Au) such that the twisted graphene surface is exposed. To ensure a reliable electrical contact during the c-AFM measurements a stencil mask was employed to deposit a 150 nm Au film contacting the graphene.

**Atomic Force Microscopy:** AFM topography and conductive AFM (c-AFM) maps were acquired on twisted graphene (TG) samples deposited on top of hBN flakes using a Cypher-S AFM (*Oxford Instruments*) in a cleanroom environment. Topography scans were acquired in AC mode, the c-AFM measurements were acquired in contact mode with a constant tip-sample potential difference of -50 mV and a scan rate 1.5 – 1.75 Hz. Values of the local twist angle were calculated from the component of the moiré wavelength along the fast scan direction to minimise the effect of temporal drift. Periodograms[54] were calculated for each scanning line and then summed along the slow scan axis to generate a magnitude spectrum in the fast direction. Identified peaks in the spectrum were then fit with gaussian functions to measure the peak frequencies, and therefore the spatial periodicity. Where the high-symmetry direction of the moiré lattice was not aligned with the fast-scan axis, we calculated the angle withstood by both directions, $\alpha$, and then applied a trigonometric transformation to obtain the period of the moiré lattice, and therefore the local twist angle. Full details are supplied in Supplementary Section 6.

**Cross-sectional STEM and EDX**: Cross sectional samples were prepared using a focused ion beam (FIB) with decreasing currents to mitigate beam damage induced during milling. Image data and elemental maps were acquired on a FEI Titan G2 80-200 S/TEM with a 4-quadrant Super-X EDX detector, operating at 200 kV accelerating voltage. The ADF STEM data was acquired with a probe current of 180 pA, a semi-convergence angle of 21.5 mrad and an annular dark field (ADF) detector inner angle of 43 mrad. EDX mapping was performed using the same probe conditions, with a pixel dwell time of 40 µs. In the map shown in figure 1d, individual x-ray counts are plotted as circles according to the colour scheme shown in the image. Both Mo and S counts are included as Mo due to peak overlap. Further details in Supplementary Section 8.

**Electroluminescence measurements** were performed at 4K, in an Attocube Attodry 1000 system. A Keithley 2614B source-measure unit was used to simultaneously bias and measure the current through the device. The bias was applied to the top graphene contact while the bottom graphene and silicon back gate were grounded. Luminescence spectra were obtained using a Princeton instruments Acton SpectraPro SP-2500 spectrometer, with a 300g/mm grating, alongside a PyLoN cryogenically cooled CCD camera.

**Competing interests:** The authors declare no competing interests.

**Data and materials availability:** Experimental data, software code for analysis and further information available upon reasonable request from the corresponding author.

**Acknowledgements:** We acknowledge support from European Union's Horizon 2020 Research and Innovation programme: European Graphene Flagship Core3 Project (grant agreement no. 881603), European Quantum Flagship Project 2DSIPC (820378), ERC Starter grant EvoluTEM (no. 715502), ERC Consolidator QTWIST (no. 101001515), ERC Consolidator Grant 3DAddChip (no. 819069), Marie Skłodowska-Curie fellowship PTMCnano (no 751883) and the Royal Society. In addition, we acknowledge support from EPSRC (grant numbers EP/V007033/1, EP/V036343/1), CDT Graphene-NOWNANO and China Scholarship Council (CSC) under grant numbers 202106950021, 201806280036 and 201908890023. This project was supported by the Henry Royce Institute for Advanced Materials, funded through EPSRC grants EP/R00661X/1, EP/S019367/1, EP/P025021/1 and EP/P025498/1. E.T. acknowledges support from OTKA Grant PD-134758 from the National Research, Development, and Innovation Office of the Ministry of Innovation and Technology of Hungary, and the Bolyai Fellowship of the Hungarian Academy of Sciences (Grant No. BO/00242/20/11).